\documentclass{PoS}

\usepackage{amsmath}    
\usepackage{amssymb} 
\usepackage{graphicx}

\def\eq#1{(\ref{#1})}
\def\s[#1\s]{\begin{align}\begin{split}#1\end{split}\end{align}}
\def\[#1\]{\begin{align}#1\end{align}}

\title{
\hfill\parbox{4cm}{ \normalsize \rm YITP-20-57}\\ \vspace{.1cm} 
Numerical and analytical studies of a matrix model with non-pairwise contracted indices}

\ShortTitle{A matrix model with non-pairwise contracted indices}

\author{\speaker{Naoki Sasakura}\\%\thanks{A footnote may follow.}\\
        Yukawa Institute for Theoretical Physics, Kyoto University, Kitashirakawa, Sakyo-ku, Kyoto 606-8502, Japan\\
        E-mail: \email{sasakura@yukawa.kyoto-u.ac.jp}}

\abstract{
The canonical tensor model, which is a tensor model in the Hamilton formalism, 
can be straightforwardly quantized and has an exactly solved physical state. 
The state is expressed by a wave function with a generalized form of the Airy function. 
The simplest observable on the state can be expressed by a matrix model which contains 
non-pairwise index contractions. This matrix model has the same form as the one that appears 
when the replica trick is applied to the spherical $p$-spin model for spin glasses,
but our case has different ranges of variables and parameters from the spin glass case.
We analyze the matrix model analytically and numerically.  We show some evidences for the presence of a continuous phase transition at the location required by a consistency condition of the canonical tensor model.
We also show that there are dimensional transitions of configurations around the transition region.
Implications of the results to the canonical tensor model are discussed.}

\FullConference{Corfu Summer Institute 2019 "School and Workshops on Elementary Particle Physics and Gravity" (CORFU2019)\\
		31 August - 25 September 2019\\
		Corfu, Greece}

\begin{document}

\section{Introduction}
Quantization of gravity is one of the most challenging problems in theoretical physics. There have been
various proposals to solve the problem. Discretized setups are used by some of them, 
including tensor models \cite{Ambjorn:1990ge,Sasakura:1990fs,Godfrey:1990dt,Gurau:2009tw}.
In discretized approaches, the main question is whether the dynamics generate global spacetimes 
with the continuum description of GR.    
In the tensor models in Euclidean forms, however, the produced objects are singular, such as branched polymers
\cite{Gurau:2011xp}, or unnatural fine-tunings of the models are required to generate global spaces 
\cite{Sasakura:2007sv,Sasakura:2007ud}. 
Currently, it is an open question whether there exist discretized theories with sensible spacetime emergence.

Stimulated by the success of CDT \cite{Loll:2019rdj}, a Lorentzian formulation of the tensor model was 
proposed by the current author \cite{Sasakura:2011sq,Sasakura:2012fb}. 
Rather than introducing an explicit time direction,  the tensor model was formulated as a first-class
constrained system in the Hamilton formalism. Namely, Hamiltonian is given by
\[
H=N_a {\cal H}_a +N_{[ab]} {\cal L}_{[ab]},
\]
where $N_a$ and $N_{[ab]}$ are arbitrary multipliers, and  
${\cal H}_a$ and ${\cal L}_{[ab]}$ are the first-class constraints satisfying the following structure of 
a closed Poisson algebra \cite{Sasakura:2011sq,Sasakura:2012fb},
\s[
\{ {\cal H},{\cal H}\}\sim {\cal L}, \\
\{ {\cal H},{\cal L}\}\sim {\cal H}, \\
\{ {\cal L},{\cal L}\}\sim {\cal L}.
\label{eq:alg}
\s]
Here the indices take the values, $1,2,\ldots,N$, and the repeated indices are assumed to be summed over.
Intriguingly, this constraint Poisson algebra has the similar form as that of the ADM formalism of GR 
with the correspondence of ${\cal H}_a $ 
and ${\cal L}_{[ab]}$ to the Hamiltonian and momentum constraints, respectively. 
In particular, the similarity includes the variable dependence of the structure function in the first line 
of \eq{eq:alg}, which makes the dynamics non-trivial. 
Under some reasonable assumptions such as time-reversal symmetry and locality for the tensor model, 
the formulation was shown to be unique \cite{Sasakura:2012fb}.
We call the tensor model the canonical tensor model for short, since this is based on the canonical formalism.

The quantization of the canonical tensor model is straightforward with the canonical quantization \cite{Sasakura:2013wza}. After quantization, the first class constraint algebra basically keeps the form \eq{eq:alg} 
with the replacement of the Poisson brackets to commutators.
The physical state conditions are given by
\[
\hat {\cal H}_a | \Psi \rangle =\hat {\cal L}_{[ab]} | \Psi \rangle =0,
\]
where  $\hat {\cal H}_a$ and $\hat {\cal L}_{[ab]}$ are the quantized constraints.
These conditions give a system of partial differential equations with non-linear coefficients for wave functions. 
Surprisingly, these equations have an exact solution \cite{Narain:2014cya}, 
\[
\Psi(P)=\left(\int_{\mathbb{R}^N} \prod_{a=1}^N d\phi_a\, \exp \left( I\, P_{abc} \phi_a \phi_b \phi_c
\right) {\rm Ai}\left(\kappa\,\phi_a \phi_a\right) \right)^{\frac{\lambda_H}{2}},
\label{eq:wavefn}
\]
where $\lambda_H=(N+2)(N+3)/2$,  $P$ designates a real symmetric tensor $P_{abc}\ (a,b,c=1,2,\ldots, N)$,
$I$ is the imaginary unit ($I^2=-1$),
and Ai$(\cdot)$ is the Airy Ai function. $\kappa$ is a real parameter, which has the opposite sign to the 
cosmological constant,
which has been argued by comparing the $N=1$ case with the mini-superspace approximation of 
GR \cite{Sasakura:2014gia}.
An important thing is that the relation $\lambda_H=(N+2)(N+3)/2$ is determined by the
hermiticity of the quantized constraint $\hat {\cal H}_a$, and should therefore have this fixed relation with $N$.

\begin{figure}[]
\begin{center}
\includegraphics[clip,width=5.0cm]{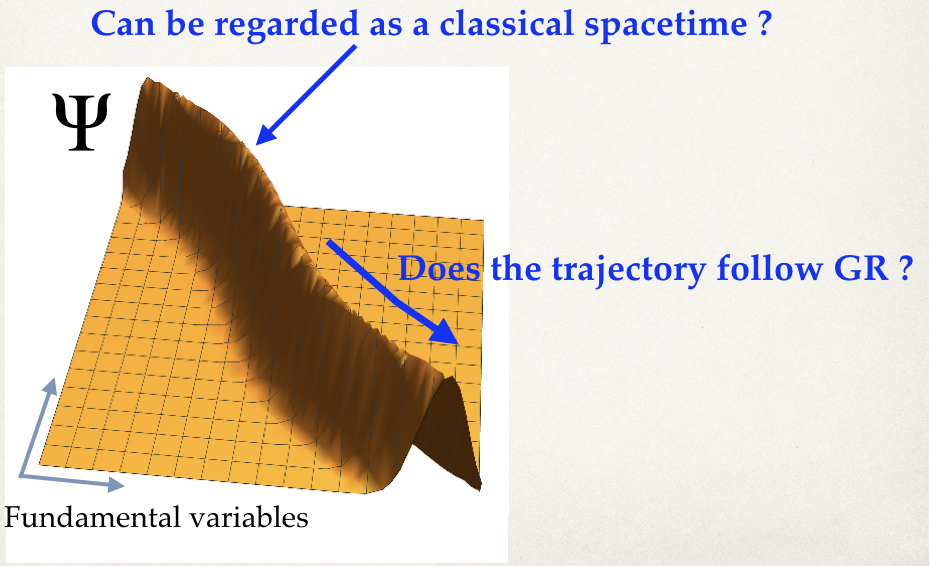}
\hfil
\includegraphics[clip,width=8.0cm]{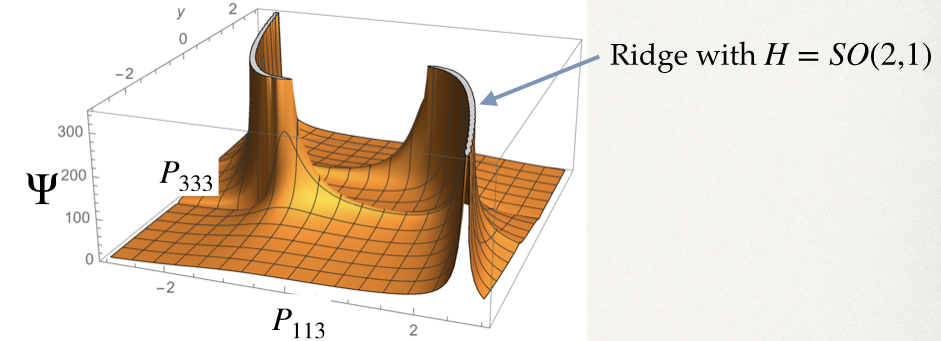}
\caption{Left: An ideal wave function which contains a ridge representing a spacetime. 
Right: The wave function of the canonical tensor model with $N=3$.  }
\label{fig:trajectory}
\end{center}
\end{figure}

The properties of the wave function \eq{eq:wavefn} are largely unknown. In particular it is unknown whether 
this has ridges which can be understood as spacetime trajectories, as in the ideal situation
shown in the left figure of Figure~\ref{fig:trajectory}. On the other hand, 
the wave function is known to have peaks (or ridges) on the locations where $P_{abc}$ satisfies 
Lie group symmetries (namely, $P_{abc}=h_{a}^{a'}h_{b}^{b'}h_{c}^{c'}P_{a'b'c'}\ (h\in H)$ with a Lie group
representation $H$),
as shown for the $N=3$ case in the right figure of Figure~\ref{fig:trajectory}  \cite{Obster:2017dhx,
Obster:2017pdq}. 
A striking fact is that the Lie-groups associated to such ridges  of the wave function \eq{eq:wavefn} 
have Lorentz signatures.   
Comparing with the left figure, the similarity encourages the study of the canonical tensor model 
as a model for emergent spacetimes.
It would also be noteworthy that Lie group symmetries, such as Lorentz, de Sitter, and gauge, are 
ubiquitous in the universe, as on the peaks (or ridges) of the wave function.  

\section{The matrix model}
\label{sec:matrix}
To understand more of the properties of the wave function, we take a different strategy in this section. 
The simplest observable would be 
\[
\langle \Psi | e^{-\alpha \hat P_{abc} \hat P_{abc}} | \Psi \rangle,
\label{eq:observable}
\]
where $\alpha$ is positive real. For example,
one of the interests in this wave function is the limit  $\alpha \rightarrow +0$.
If the wave function contains a ridge which has a spacetime interpretation
with an infinite extension of a time direction, the norm will linearly diverge in this time direction
in the $\alpha\rightarrow +0$ limit. 
Therefore the behavior of \eq{eq:observable} in the limit is related to whether the tensor model wave 
function contains an infinite extension of time direction or not.

Currently, it is not easy to study \eq{eq:observable}. Therefore our strategy is to carry out some simplifications
to the wave function \eq{eq:wavefn}. When $\kappa>0$, which corresponds to the negative cosmological
constant, the Airy Ai function is a fast decaying function of $\phi_a \phi_a$. 
Therefore, as an initial study, it would be interesting to simplify the Airy function to $e^{-k \phi_a\phi_a}$ 
with positive $k$, and study the properties of \eq{eq:observable}.
With this simplification we obtain
\[
\langle \Psi_{simple} | e^{-\alpha \hat P_{abc} \hat P_{abc}} | \Psi_{simple}\rangle=
{\cal N} \alpha^{-\# P/2}Z_{N,\lambda_H}((4 \alpha)^{-1},k),
\label{eq:reltensor}
\]
where ${\cal N}$ is an unimportant normalization factor, $\# P=N(N+1)(N+2)/6$, 
namely, the number of independent components of $P_{abc}$,
and $Z_{N,R}(\lambda,k)$ is the partition function 
of the matrix model defined by \cite{Lionni:2019rty}
\[
Z_{N,R}(\lambda,k)=
\int \prod_{i=1}^R \prod_{a=1}^N d\phi_a^i \ e^{-\lambda \sum_{i,j=1}^R (\phi_a^i\phi_a^j)^3-k\sum_{i=1}^R \phi_a^i\phi_a^i}.
\label{eq:matrix}
\]
Here the repeated lower indices are assumed to be summed over, but the summation over the upper indices 
are written explicitly. This is because the upper indices are not always contracted pairwise:
the first term of the exponent contains triple contractions of the upper indices.
We also assume $\lambda,k>0$ for the obvious convergence of the integration.

It is particularly important to note that, while the matrix model \eq{eq:matrix} itself has
both $N$ and $R$ as free 
parameters, they are restricted by $R=\lambda_H=(N+2)(N+3)/2$, if we consider 
the relation \eq{eq:reltensor} to the tensor model.
Surprisingly, we will see later that this relation agrees with the location of the continuous phase transition 
(or crossover) point
in the leading order of $N$.
 
 Lastly, let us comment on the relation between our matrix model and the one that appears in the replica 
trick applied to the the spherical $p$-spin model ($p=3$) for spin glasses \cite{pspin,pedestrians}. 
The partition function of the spherical $p$-spin model for $p=3$ is given by
\[
Z_{p\hbox{-} {\rm spin}}=\int_{\phi_a \phi_a=1}  d\phi \ e^{-P_{abc} \phi_a \phi_b \phi_c}
\]
with a random coupling $P_{abc}$ to simulate spin glasses. Simulating the random coupling by
Gaussian distribution, $e^{-\alpha P^2}$, and applying the replica trick, one obtains
\[
\int dP\, e^{-\alpha P^2} (Z_{p\hbox{-}{\rm spin}})^R=\int_{\phi_a^i\phi_a^i=1} d\phi\ e^{\frac{1}{4 \alpha} \sum_{i,j=1}^R 
(\phi_a^i\phi_a^j)^3}.
\] 
This is very similar to our matrix model \eq{eq:matrix}, but there are a few major differences. 
One is that the coupling parameter has the opposite sign compared to our case. 
This means that the dominant configurations 
are largely different between the spin glass and our cases. 
Another difference is that, in the replica trick, the replica number $R$ is finally taken to 
the limit $R\rightarrow 0$, while we are rather interested in the case of the tensor model $R=\lambda_H=
(N+2)(N+3)/2$. In particular, this means that, in the thermodynamic limit $N\rightarrow \infty$,
$R$ should also be taken infinite for the correspondence to the tensor model. 

\section{Perturbative computations}
\begin{figure}[]
\begin{center}
\includegraphics[clip,width=10.0cm]{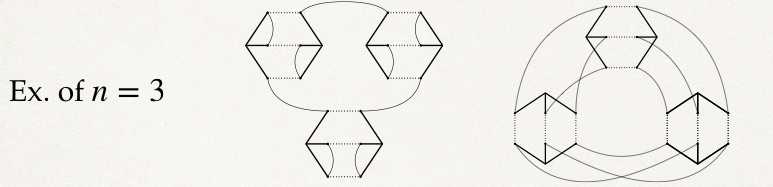}
\caption{The necklace diagrams for the third order.}
\label{fig:necklace}
\end{center}
\end{figure}

A perturbative computation of the matrix model \eq{eq:matrix} was carried out in \cite{Lionni:2019rty}. 
Regarding the exponent of \eq{eq:matrix} as the action of the matrix model, Feynman rules can be
written down for the perturbative computation in $\lambda$. 
The most dominant diagrams in large $R$ for each order in $\lambda$ are the necklace 
diagrams depicted in Figure~\ref{fig:necklace}, where the examples of the order $n=3$ are given. 
Summing over all $n$, it has been obtained that 
\[
Z_{N,R}(\lambda,k)={\cal N'} \int dr\, r^{NR-1} f_{N,R}(\lambda r^6)\, e^{-k r^2}, 
\]
where 
\[
f_{N,R}(t)=\left(1+\frac{12 t }{N^3 R^2}\right)^{-\frac{ N(N-1)(N+4)}{12}}
\left(1+\frac{6(N+4) t }{N^3 R^2}\right)^{-\frac{N}{2}}.
\label{eq:f}
\]

A generalization of the perturbative computation above was performed in \cite{Sasakura:2019hql}
by employing a slightly different strategy. 
In the generalization, the coupling term in the exponent was generalized to 
\[
\sum_{i,j=1}^R 
\Lambda_{ij} (\phi_a^i\phi_a^j)^3
\]   
with a symmetric matrix $\Lambda$.
In this case, the above $f_{N,R}(t)$ is generalized to 
\[ 
f_{N,R}(t)=\prod_{e_{\Lambda}} h_{N,R}(e_{\Lambda} t),
\label{eq:fleadinggen}
\]
where the product is over the eigenvalues of the matrix $\Lambda$ with the degeneracies taken
into account, and 
\[
h_{N,R}(t)=(1+12 \gamma_3 t)^{-\frac{N(N+4)(N-1)}{12}} \, (1+6 (N+4) \gamma_3 t)^{-\frac{N}{2}} 
\]
with 
\[
\gamma_3=\frac{\Gamma\left(\frac{NR}{2}\right)}{8\, \Gamma\left( \frac{NR}{2}+3\right)}.
\]
This indeed agrees with \eq{eq:f} for the special case $\Lambda_{ij}=\lambda$
in the leading order of $R$, since the eigenvalues of $\Lambda$ are one $R\lambda$ and $R-1$ zeros.
This generalization makes it possible to compute the expectation values of the products of $(\phi_a^i\phi_a^j)^3$
by taking the derivatives of the partition function with respect to $\Lambda_{ij}$.

\section{Monte Carlo simulations}
The Monte Carlo simulations were performed by the Metropolis update method in \cite{Sasakura:2019hql}.
We observed a phase transition or a crossover around $R\sim N^2/2$. This location curiously agrees 
with $R=\lambda_H=(N+2)(N+3)/2$  in the leading order of $N$,
which is required by the consistency of the tensor model, as explained in Section~\ref{sec:matrix}.

Since the overall coefficient of the exponent of \eq{eq:matrix} can be absorbed by the rescaling of 
$\phi_a^i$, we set $\lambda=1$ in the simulations, leaving $N,R,k$ as variables.

\subsection{Expectation values of observables}
\begin{figure}[]
\begin{center}
\includegraphics[clip,width=4.0cm]{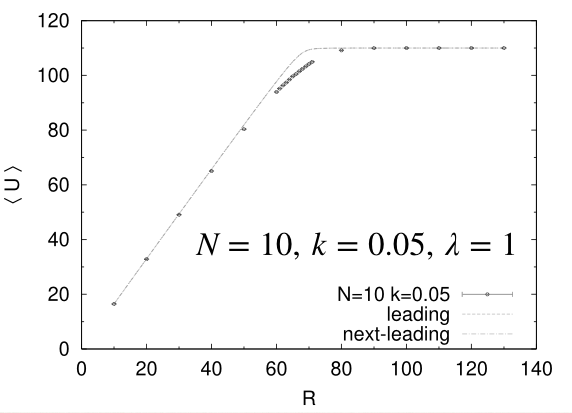}
\hfil
\includegraphics[clip,width=4.0cm]{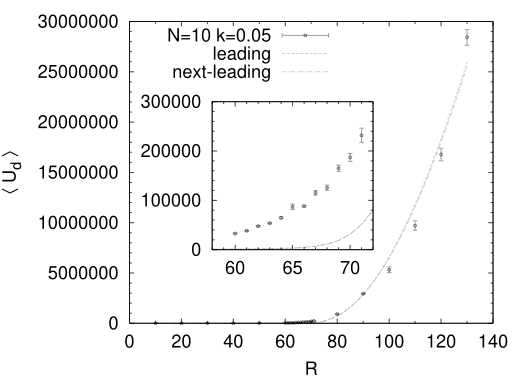}
\hfil
\includegraphics[clip,width=4.0cm]{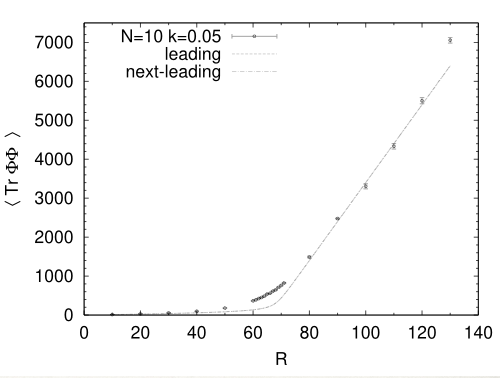}
\caption{Expectation values of observables. The dots with error bars are the Monte Carlo results. The solid
lines are the perturbative analytical results. }
\label{fig:exp}
\end{center}
\end{figure}

We show the results of the expectation values, $\langle U \rangle$, $\langle U_d \rangle$, and $\langle r^2 \rangle$,  in Figure~\ref{fig:exp}, where
\[
U=\sum_{i,j=1}^{R} (\phi_a^i \phi_a^j)^3,\ 
U_d=\sum_{i=1}^R (\phi_a^i \phi_a^i)^3,\ 
r^2=\sum_{i=1}^R \phi_a^i \phi_a^i.
\]
There seems to exists a continuous phase transition point or a cross over near $R\sim  N^2/2$. 
The simplest characterization of the transition is that $\phi_a^i$ take small values for $R\lesssim N^2/2$,
while $\phi_a^i$ take large values for $R\gtrsim N^2/2$. 
To answer the question whether this is a phase transition in the thermodynamic limit $N\rightarrow \infty$ 
turned out to be very difficult.
In the region where $R\gtrsim N^2/2$ and $k$ is small, there are slowdowns of Monte Carlo updates:
If we took step sizes so that acceptance rates were reasonable, the configuration updates were too slow
for us to see the thermodynamic equilibriums within our runtimes. 
This problem restricted us to perform the simulations
in the range $N\lesssim 10$ and $k\gtrsim 0.05$, which were not enough to see a sharp transition:
The transition may be a phase transition or just a crossover. 

Another issue is that the analytical results do not agree with the Monte Carlo results 
in the vicinity of the transition region $R\sim N^2/2$. 
This implies that some important contributions have not been 
taken into account in the perturbative analytic computations.

\subsection{Dimensional transition}
To see what occurs at the transition region in more detail, we studied the characteristics of the configurations.
The $N$-dimensional vector $\phi_a^i$ for each $i$ can be regarded as representing a point in an 
$N$-dimensional vector space. Then the collection of $R$ points, $\phi_a^i\ (i=1,2,\ldots,R)$, represents
a point cloud in the $N$-dimensional space.  Rather than considering the point cloud itself, we
project the points onto the unit sphere $S^{N-1}$ by
$\tilde \phi_a^i=\phi_a^i/\sqrt{\phi_a^i \phi_a^i}$.
If a point cloud is a $d+1$ dimensional ball-like object
in the $N$-dimensional space, the projection produces a point cloud with the topology of $S^{d}$ 
on the unit $S^{N-1}$. Therefore the topology of the projected point cloud on $S^{N-1}$
can be used to determine the dimension of the point cloud of $\phi_a^i\ (i=1,2,\ldots,R)$.
One can know the topology in various manners. Below we will study the topologies of the
projected point clouds  in three different manners, where the datas
come from the actual simulations.

\begin{figure}[]
\begin{center}
\includegraphics[clip,width=10.0cm]{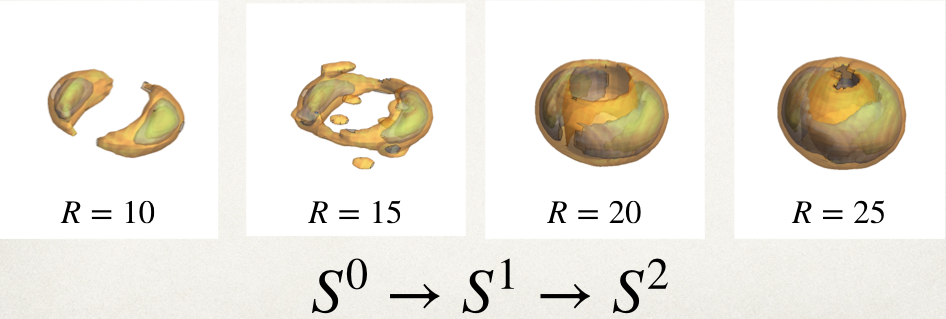}
\caption{The density plots of the projected point clouds, which were projected further to 
a three-dimensional space by PCA. The datas are from the simulations of $N=4,\ k=0.01$ and 
the values of $R$ shown in the figure.  }
\label{fig:PCA}
\end{center}
\end{figure}

First, we use Principal Component Analysis (PCA). 
This is to extract some major directions in which a point cloud has extensions. If $d\leq 2$, 
one can project $S^d$ into a three-dimensional space as shown in Figure~\ref{fig:PCA}.
There exist topology changes from lower to higher dimensional spheres,
when $R$ is increased in the transition region. 
 
\begin{figure}[]
\begin{center}
\includegraphics[clip,width=10.0cm]{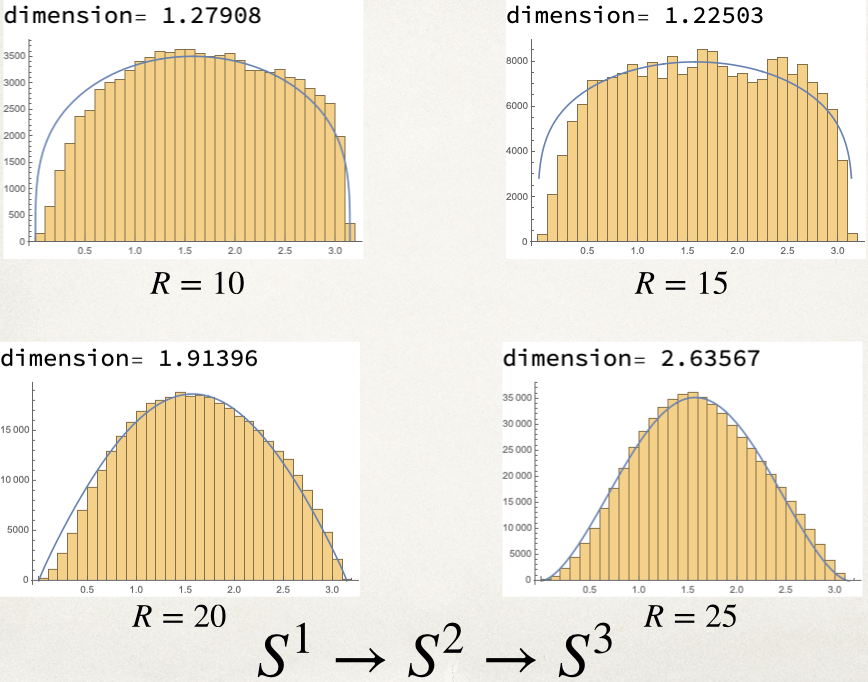}
\caption{The angular distribution among $\phi_a^i$ and $\phi_a^j\ (i\neq j)$ for various $R$ with 
$N=10,\ k=0.01$.}
\label{fig:angular}
\end{center}
\end{figure}

Next we will see the topology by the angular distribution among the vectors, $\tilde \phi_a^i$ and 
$\tilde \phi_a^j\ (i\neq j)$. It is easy to understand that,  if $\tilde \phi_a^i\ (i=1,2,\ldots,R)$
form a point cloud of the form of a unit $S^{d}$, the angular distribution will be 
\[
{\cal N} \sin^{d-1}(\theta) d\theta
\label{eq:angular}
\]      
with an unimportant normalization factor ${\cal N}$. We fit the data with the distribution \eq{eq:angular}
to extract the value of $d$.
The results are shown in Figure~\ref{fig:angular}. There is an increase of the dimensions with the increase
of $R$.

\begin{figure}[]
\begin{center}
\includegraphics[clip,width=7.6cm]{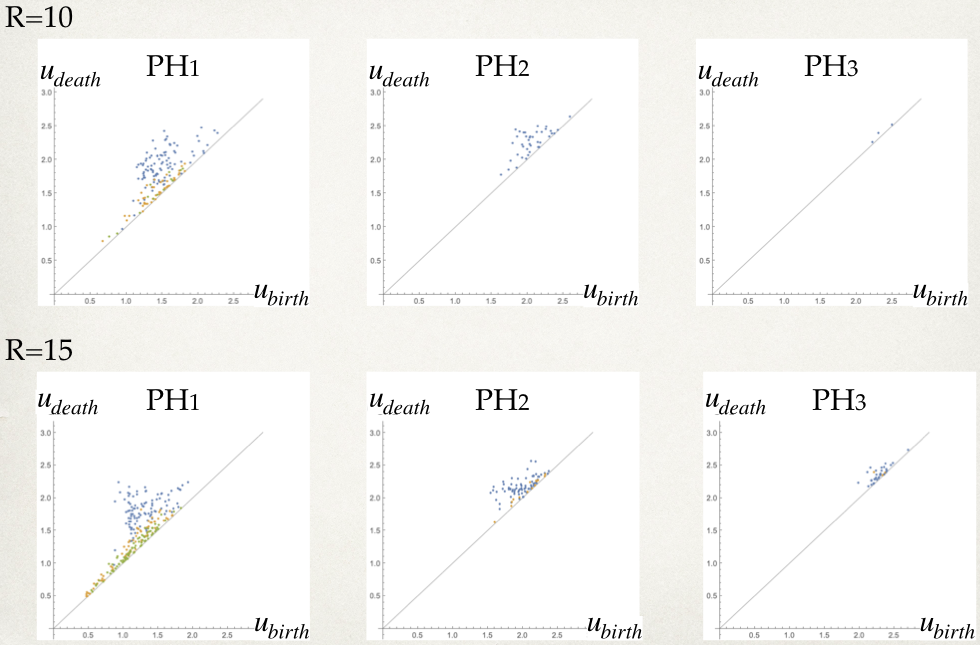}
\hfil
\includegraphics[clip,width=7.2cm]{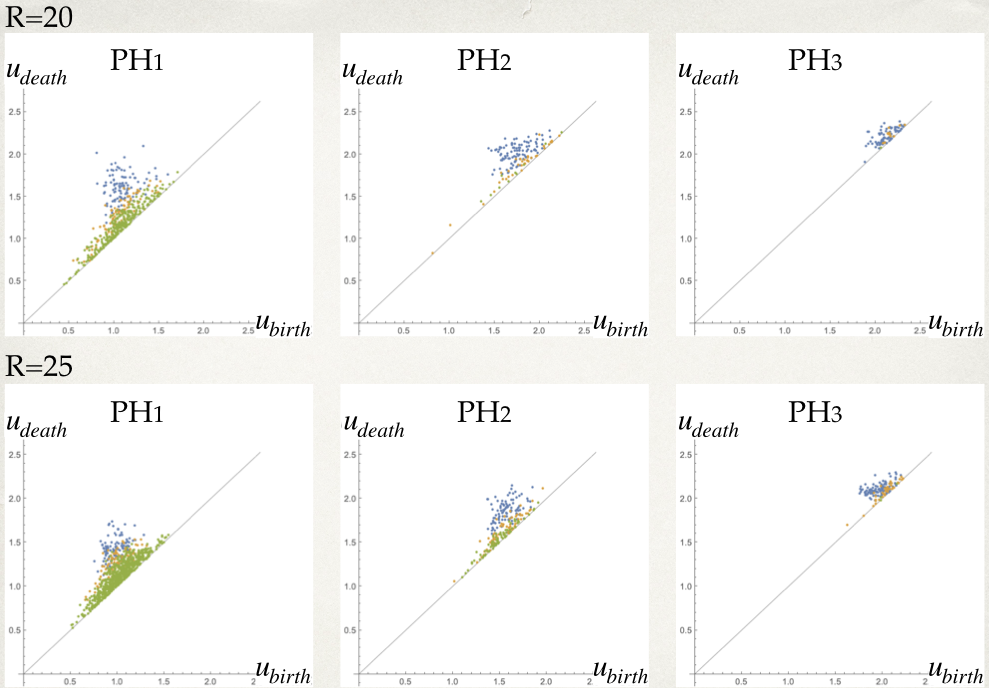}
\caption{Persistent diagrams for various $R$ with $N=10,\ k=0.01$, where persistent homology group 
elements are represented by dots.
The scale parameters, $u_{birth}$ and $u_{death}$, designate the births and deaths of the elements. 
Long-life (persistent) elements are the dots distant from the diagonal lines.}
\label{fig:persistent}
\end{center}
\end{figure}

The last way is to use the persistent homology \cite{carlsson_topology_2009}. 
This is a method used in the topological data analysis, that
extracts homology structure of datas which are coarse-grained by a scale parameter. 
If a point cloud forms $S^{d}$, 
a $d$-dimensional persistent homology group element will be detected. 
The result of the analysis is shown in Figure~\ref{fig:persistent}, where $PH_i$ designates
the $i$-dimensional persistent homology group. As can be seen, higher dimensional homology group
elements appear more often as $R$ is increased.

\section{Summary and future prospects}
We have studied a matrix model which represents the simplest observable for an exactly solved physical state of 
the canonical tensor model. This matrix model has the unusual property that it contains non-pairwise 
index contractions, and cannot be exactly solved by the well-known methods applied to the usual
matrix models. This matrix model has the similar form as the one that appears in the replica trick of the 
spherical $p$-spin model ($p=3$) for spin glasses. However, our case has different ranges of parameters 
and variables from the spin glass case, and new analysis is needed. 
We have performed some analytical and numerical computations for the matrix model.
The most striking result is the presence of the continuous phase transition point
or a crossover at the location which is required by the consistency of the canonical tensor model. 
This means that the canonical tensor model exists exactly on or in the vicinity of the 
continuous phase transition point. 
This means that the canonical tensor model automatically exists at the location 
where its continuum limit could be taken,
as continuum theories are often obtained by taking continuum limits around continuous phase transition points.   
The non-trivial behavior of the dimensions of the configurations, represented by point clouds,
would also be encouraging for the possibility of spacetime emergence in the canonical tensor model.    

Further improvement of the results after the conference has been presented in our recent 
paper \cite{Obster:2020vfo}.
  
\vspace{1cm}
\section*{Acknowledgements}
%%%%%%%%%%%%%%%%%%%%%%%%%%%%%%%%%%%%%%%%%%%%%%%
%\centerline{\bf Acknowledgements} 
The Monte Carlo simulations in this study were performed using KEKCC, the cluster system of KEK.
N.S. would like to thank L.~Lionni and S.~Takeuchi for the collaborations, on which this presentation is based.
The work of N.S. is supported in part by JSPS KAKENHI Grant No.19K03825. 
 %%%%%%%%%%%%%%%%%%%%%%%%%%%%%%%%%%%%%%%%%%%%%%%%

\end{document}